\documentclass[checkin,showpacs,psfig,aps,prb]{revtex4}

\usepackage{mathrsfs}

\usepackage{color}

\newcommand{\ub}{\mu_{\rm B}}
\newcommand{\bn}{\begin{enumerate}}
\newcommand{\en}{\end{enumerate}}
\newcommand{\ba}{\begin{eqnarray}}
\newcommand{\ea}{\end{eqnarray}}

\newcommand{\mn}{Mn$_2$RuGa }
\newcommand{\mrg}{Mn$_2$RuGa }
\newcommand{\mrge}{Mn$_2$RuGa}

\usepackage{graphicx,color}

\newcommand{\dz}{$d_{z^2}$ }

\newcommand{\be}{\begin{equation}}
\newcommand{\ee}{\end{equation}}
\newcommand{\la}{\langle}

\newcommand{\ra}{\rangle}
\newcommand{\et}{{\it et al. }}

\def\prl{{ Phys. Rev. Lett. }}

\begin{document}

\newcommand{\clr}{}
\newcommand{\clra}{}











\title{First-principles insights into  all-optical spin
  switching in the half-metallic Heusler ferrimagnet Mn$_2$RuGa}



\author{G. P. Zhang$^*$}

 \affiliation{Department of Physics, Indiana State University,
   Terre Haute, IN 47809, USA }

\author{Y. H. Bai}

\affiliation{Office of Information Technology, Indiana State
  University, Terre Haute, IN 47809, USA }

\author{M. S. Si}

\affiliation{School of Materials and Engineering, Lanzhou University,
  Lanzhou 730000, China}

 \author{Thomas F. George} \affiliation{Departments of Chemistry \&
   Biochemistry and Physics \& Astronomy \\University of
   Missouri-St. Louis, St.  Louis, MO 63121, USA }

\date{\today}

\begin{abstract}
  {All-optical spin switching (AOS) represents a new frontier in
    magnetic storage technology -- spin manipulation without a
    magnetic field, -- but its underlying working principle is not
    well understood. Many AOS ferrimagnets such as GdFeCo are
    amorphous and renders the high-level first-principles study
    unfeasible.  The crystalline half-metallic Heusler \mrg presents
    an opportunity.  Here we carry out hitherto the comprehensive
    density functional investigation into the material properties of
    \mrge, and introduce two concepts - the spin anchor site and the
    optical active site - as two pillars for AOS in ferrimagnets. In
    \mrge, Mn$(4a)$ serves as the spin anchor site, whose band
    structure is below the Fermi level and has a strong spin moment,
    while Mn$(4c)$ is the optical active site whose band crosses the
    Fermi level. Our magneto-optical Kerr spectrum and band structure
    calculation jointly reveal that the delicate competition between
    the Ru-$4d$ and Ga-$4p$ states is responsible for the creation of
    these two sites. These two sites found here not only present a
    unified picture for both \mrg and GdFeCo, but also open the door
    for the future applications. Specifically, we propose a
    Mn$_2$Ru$_x$Ga-based magnetic tunnel junction where a single laser
    pulse can control magnetoresistance.}
\end{abstract}




 \maketitle


\section{Introduction}

Laser-induced ultrafast demagnetization \cite{eric} changes the
landscape of spin manipulation, where the laser field plays a central
role in magnetism.  All-optical spin switching (AOS)
\cite{stanciu2007} is a prime example, where a single laser pulse can
turn spins from one direction to another, free of an external magnetic
field.  As more and more materials are discovered
\cite{mangin2014,lambert2014,vomir2017}, a critical question on the
horizon is what properties are essential to AOS. Earlier studies have
focused on magnetic orderings such as ferrimagnetic versus
ferromagnetic \cite{schubert2014a}, sample composition
\cite{hassdenteufel2015}, compensation temperature
\cite{vahaplar2012}, magnetic domains \cite{elhadri2016b} and others
\cite{ourbook}, but most AOS materials are amorphous and difficult to
simulate within state-of-the-art density functional theory.  This
greatly hampers the current effort to decipher the mystery of AOS at a
microscopical level that goes beyond the existing phenomenological
understanding \cite{ostler2012}.

Heusler compounds represent a new opportunity
\cite{galanakis2002b,wollmann2017}. Their properties can be
systematically tailored, only subject to the structure
stability. Different from rare-earth-transition metals
\cite{stanciu2007,hassdenteufel2013}, one has an empirical
Slater-Pauling rule to predict spin moments
\cite{wurmehl2005,wurmehl2006,liu2006,wollmann2015,song2017,suzuki2018,yang2015,zhang2018a}.
Although this rule is simple \cite{leuken1995}, the actual synthesis
of a desired material is a monumental task of decades in making
\cite{rode2013}, because many materials are unstable experimentally.
In 2002, Hori \et \cite{hori2002} successfully synthesized various
(Mn$_{1-x}$Ru$_x$)$_3$Ga alloys with $x=0.33-0.67$ and determined the
spin moment of 1.15 $\ub$ per formula.  In 2014, Kurt \et
\cite{kurt2014} demonstrated that Ru can significantly reduce the spin
moment in ferrimagnet Mn$_2$Ru$_x$Ga. Because one can tune composition
$x$, Mn$_2$Ru$_x$Ga is likely to be a half-metal and fully-compensated
ferrimagnet \cite{nayak2015,sahoo2016}, with no stray field, ideal for
spintronics \cite{betto2016,wollmann2017,finley2020}.  Research has
intensified immediately
\cite{galanakis2014,yang2015,zic2016,fleischer2018,siewierska2021,chatterjee2021}.
Lenne \et \cite{lenne2019} found that the spin-orbit torque reaches
$10^{-11}$Tm$^2$/A in the low-current limit.  Banerjee \et
\cite{banerjee2020} reported that a single 200-fs/800-nm laser pulse
can toggle the spin from one direction to another in \mn within 2 ps
or less. Just as found in GdFeCo \cite{stanciu2007,ostler2012}, for
every consecutive pulse, the spin direction is switched.  This
discovery \cite{banerjee2020} demonstrates the extraordinary
tunability of Heusler compounds, which now changes the trajectory of
AOS research \cite{aip20,banerjee2021,jakobs2021}, with \mn as a
crystalline model system where the first-principles investigation is
now possible.

In this paper, we carry out the comprehensive first-principles
density-functional study to pin down the material properties essential
to all-optical-spin switching in ferrimagnet \mrge.  We introduce two
concepts - the spin anchor site (SAS) and the optical active site
(OAS) as two essential pillars of AOS in ferrimagnets.  SAS has a
strong spin moment, and in \mrg it is the Mn$(4a)$ site. Our band
structure reveals that the Mn$(4a)$'s band is 0.5 eV below the Fermi
level. By contrast, OAS has a smaller spin moment, easier to be
switched optically \cite{epl16}, and in \mrg it is the Mn$(4c)$
site. Its band is around the Fermi level and accessible to optical
excitation \cite{bonfiglio2021}. The creation of SAS and OAS is the
making of Ru and Ga. The Ru-$4d$ electrons set up the initial spin
configuration with a strong spin moment concentrated on the distant
Mn$(4c)$, but Ga tips this balance and reverses the relative spin
magnitude between Mn$(4a)$ and Mn$(4c)$. Although Ru and Ga are weakly
magnetic, their energy bands appear in the same energy window as two
Mn atoms, which is manifested in the magneto-optical Kerr
spectrum. Guided by two essential sites, we can now unify \mrg with
GdFeCo, despite of their apparent structural differences, and extract
three essential properties for AOS. (i) A ferrimagnet must have a spin
anchor site, i.e. Gd in GdFeCo and Mn$(4a)$ in \mrge. (ii) It must
have an optical active site, i.e.  Fe in GdFeCo and Mn$_2(4c)$ in
\mrge. (iii) Its spin anchor site and optical active site must be
antiferromagnetically coupled to minimize the potential energy barrier
\cite{jap19}. We propose a laser-activated magnetic tunnel junction
based on the same material Mn$_2$Ru$_x$Ga, but with different
compositions $x$ which form optical activation, spin filtering and
reference layers.  This device, if successful, represents an ideal
integration of fully-compensated half-metallicity in spintronics into
all-optical spin switching in femtomagnetism
\cite{ourreview,rasingreview}.

{\clr
The rest of the paper is arranged as follows. In Sec. II, we present
our theoretical formalism.   Section III is devoted to the results and
discussion, which includes the crystal structure, electronic band
structure, ultrafast demagnetization, and Kerr rotation
angle. Finally, we conclude this paper in Sec. IV. 
}

\newcommand{\nk}{n{\bf k}}
\newcommand{\mk}{m{\bf k}}

\section{Theoretical formalism and calculation} 

Element Mn lies in the middle of $3d$ transition metals, with a
half-filled $3d$ shell and zero orbital moment, just as Gd in the
middle of $4f$ rare-earth metals.  Mn is the only $3d$ transition
metal element in inverse Heusler compounds, which is similar to a
rare-earth element \cite{wollmann2014}.  \mn crystallizes in an
inverse $XA$ Heusler structure
\cite{wollmann2014,galanakis2014,yang2015,zhang2018a,chatterjee2021}
(see Fig. \ref{fig01}(a)), where two manganese atoms, Mn$_1$ and
Mn$_2$, are situated at two distinct Wyckoff positions $4a(0,0,0)$ and
$4c(\frac{1}{4},\frac{1}{4},\frac{1}{4})$, and are
antiferromagnetically coupled.  Ru and Ga sit at
$4d(\frac{3}{4},\frac{3}{4},\frac{3}{4})$ and
$4b(\frac{1}{2},\frac{1}{2},\frac{1}{2})$, respectively. The inverse
Heusler $XA$ structure has two Mn atoms separated by a vector
$(\frac{1}{4}, \frac{1}{4}, \frac{1}{4})$, while the $L2_1$ structure
by $(\frac{1}{2}, \frac{1}{2}, \frac{1}{2})$
\cite{kurt2014,zic2016,fleischer2018}.  The experimental lattice
constants in this nearly cubic material are $a=b=c=5.97 \rm \AA$
\cite{kurt2014}.  Viewing along the diagonal direction, four atoms
form chains $\rm Mn_1$-$\rm Mn_2$-Ga-Ru-$\rm Mn_1 \cdots$.  Therefore,
\mrg loses both inversion and time reversal symmetries due to the
antiferromagnetic coupling between two Mn atoms.

We employ the state-of-the art density functional theory and the
full-potential linearlized augmented plane wave (FLAPW), as
implemented in the Wien2k code \cite{wien2k}.  We first
self-consistently solve the Kohn-Sham equation\be \left
[-\frac{\hbar^2\nabla^2}{2m_e}+V_{Ne}+V_{H}+V_{xc} \right
]\psi_{\nk}(r)=E_{\nk} \psi_{\nk} (r), \label{ks} \ee where
$\psi_{\nk}(r)$ is the wavefunction of band $n$ at the crystal
momentum ${\bf k}$ and $E_{\nk}$ is its band energy.  The terms on the
left are the kinetic energy operator, the attraction between the
nuclei and electrons, the Hartree term, and the exchange-correlation
\cite{pbe}, respectively.  The spin-orbit coupling is included using a
second-variational method in the same self-consistent iteration.

\section{Results and discussions}

{\clr

\subsection{Crystal structure}

Ordered Heusler alloys have three distinctive kinds of
structures \cite{hakimi2013}: (1) normal full-Heusler $X_2YZ$ alloys
with group symmetry $L2_1$, (2) half Heusler $XYZ$ compounds with
group symmetry $C1_b$, and (3) inverse Heusler $X_2YZ$ alloys with
group symmetry $XA$. (1) has the space group No. 225.  (2) and (3)
have the same space group No. 216.  $L2_1$ has $(8c)$ site, which is
split into two different sites in $XA$ \cite{wollmann2014}.  However,
over the years, various Wyckoff positions are adopted in the
literature.  For the $XA$ structure, Wollmann \et \cite{wollmann2014}
used a different set of Wyckoff positions for Mn at
$4d(\frac{1}{4},\frac{1}{4},\frac{1}{4})$, $Y$ at $4c(
\frac{3}{4},\frac{3}{4},\frac{3}{4})$, Mn at
$4b(\frac{1}{2},\frac{1}{2},\frac{1}{2})$, and $Z$ at $4a(0,0,0)$. So
in their paper, their $(4a),(4b),(4c),(4d)$ positions have different
meanings from those in \cite{galanakis2014}.  In order to convert
Wollmann's notation to the latter notation, one has to shift the
entire cell by $4d(\frac{1}{4},\frac{1}{4},\frac{1}{4})$.  For the
$L2_1$ structure, Wollmann \et \cite{wollmann2014} also adopted
different positions, which were again used in their review paper
\cite{wollmann2017} (see Table \ref{table0}).

In Mn$_2$RuGa, several versions have also been used. It adopts an $XA$
structure. Kurt \et \cite{kurt2014} correctly assigned the space group
symmetry $L2_1$ to the full Heusler compound, but inappropriately
assigned the same group to Mn$_2$RuGa, and so did Zic \et
\cite{zic2016}. Both Zic \et \cite{zic2016} and Fleischer \et
\cite{fleischer2018} had the correct notations for all the atoms, but
their figure switched the positions for Ru and Mn$_2$, where Ru$(4d)$
appears at position $4c(\frac{1}{4},\frac{1}{4},\frac{1}{4})$ and
Mn$_2(4c)$ at $4d(\frac{3}{4},\frac{3}{4},\frac{3}{4})$.  Since they
never used the figure to characterize their experimental data, this
change does not affect their results.  We also notice that Betto \et
\cite{betto2015} assigned $C1_b$ group symmetry to \mrge, where two Mn
atoms are at $4a(0,0,0)$ and $4c(3/4,3/4,3/4)$ while Ru at
$4d(1/4,1/4,1/4)$ and Ga at $4b(1/2,1/2,1/2)$. One can see from Table
\ref{table0} that $C1_b$ has no $4c$ site.

Galanakis \et \cite{galanakis2014} exchanged the positions for Ru and
Ga, so Ru is at site $4b$ and Ga is at site $4d$.  Although in general
such an exchange is allowed, they do not match the existing
experimental results \cite{bonfiglio2021}.  For instance, only
Mn$_1(4a)$ has Ru as its neighbor. If we exchange the positions for
Ru and Ga, then Mn$_2(4c)$ would have Ru as its neighbor.

We summarize those used Wyckoff positions in the same table, so the
reader can see the difference.  We adopt the common convention, as
listed in the last line in Table \ref{table0}. This convention matches
the experimental results better \cite{fleischer2018}. In particular,
the magneto-optics signal agrees with the experimental one.

}

\subsection{Band structure}

We choose a big $k$ mesh of $44\times 44\times 44$, with 11166
irreducible $k$ points in the Brillouin zone. The product of the
Muffin-tin radius $R_{\rm MT}$ and the planewave cutoff is 7, where
$R_{\rm MT}(\rm Mn_1,Mn_2, Ru)=2.42$ bohr, and $R_{\rm MT}(\rm
Ga)=2.28$ bohr.  We find that Mn$_1$ has spin moment of
$M_{4a}=3.17\ub$. We call Mn$_1(4a)$ the spin anchor site, SAS, as it
pins the magnetic configuration, so the magnetic structure can be
stabilized and is immune to optical excitation.  Mn$_2(4c)$ atoms form
another spin sublattice with a smaller spin moment of
$-M_{4c}=-2.31\ub$.  The entire cell has the spin moment of 1.027
$\ub$, in agreement with prior studies \cite{galanakis2014,yang2015}.
Figure \ref{fig01}(a) shows the spatial valence spin density
integrated from 2 eV below the Fermi level for each atom, where the
red(blue) color refers to the majority(minority) spin. One can see the
spin density is mainly localized on these two Mn atoms, where Mn$_1$
has a larger spin in the spin up channel and Mn$_2$ has the spin
density in the spin down channel, so they are antiferromagnetically
coupled.

Our first finding is that the above spin configuration hinges on the
delicate balance between Ru and Ga. Figure \ref{fig01}(b) shows their
respective atomic energies.  Ru's $4d^75s^1$ states are close to Mn's
unoccupied $5d^0$ states.  Without Ga, when Mn and Ru form a solid,
the spin moment on Ru increases by five times to 0.39$\mu_B$ and is
antiferromagnetically coupled to the Mn$_1(4a)$'s spin, but its spin
is now ferromagnetically coupled to the distant Mn$_2(4c)$, which is
opposite to the native \mrg (see Table \ref{tab}). Adding Ga tips the
balance, because Ga's $4p^1$ is higher than Mn's unoccupied $5d^0$
orbitals (see Fig. \ref{fig01}(b)), so Ga can transfer electrons to Mn
atoms more easily than Ru. We integrate the atom-resolved density of
states around each sphere, and find that the number of the Ru $4d$
electrons is 5.87, reduced by 1.13 with respect to its atomic $4d^7$,
while the $4p$ electron of Ga is 0.81, reduced by 0.2 from $4p^1$.
The total number of electrons within the Mn$_1(4a)$ and Mn$_2(4c)$
spheres are almost exactly the same, 6.04, but the number of $3d$
electrons in each spin channel is very different. Table \ref{tab}
shows that Mn$_1(4a)$ has 4.09 $3d$ electrons in the majority channel
and 1.01 in the minority channel, in contrast to Mn$_2(4c)$ where
1.44 and 3.72 electrons are present. The total number of $3d$
electron is still close to 5.  Table \ref{tab} summarizes these
results. In general, the orbital moment on Mn$_1$ is small, around
0.025$\ub$, and that on Mn$_2 $ is slightly larger, reaching
-0.046$\ub$, which is beneficial to the spin-orbit torque
\cite{jap19,aip20}, important for AOS \cite{banerjee2020}.

Figure \ref{fig2}(a) shows the band structure, superimposed with the
Mn$_1$-$3d$ orbital character from its spin majority channel.  The
orbital characters are highlighted by the circles, whose radius is
proportional to the weight of the Mn$_1$-$3d$ character, and the lines
are the actual band dispersion.  Bands with a clear dominance of a
single orbital are highlighted, and in the figure, \dz and
$d_{x^2-y^2}$ are denoted by $z^2$ and $x^2-y^2$ for simplicity; and
this is the same for other orbitals. The entire set of detailed
orbital characterization is presented in \cite{sm}.  We see that the
Mn$_1$'s occupied majority band centers around -0.6 eV below the Fermi
level $E_F$ (horizontal dashed line), with a smaller contribution
close to the Fermi level.  This feature is reflected in the
$3d$-partial density of states (pDOS) in Fig. \ref{fig2}(b), where a
small peak at the Fermi level is found, consistent with two prior
studies \cite{galanakis2014,yang2015}, indicative of structural
instability \cite{wollmann2014}.  Figure \ref{fig2}(c) shows that the
Mn$_1$ spin minority band has a single $d_{xz}/d_{yz}$ band, which
crosses the Fermi level from the L to $\Gamma$, and then to X point,
but this single band crossing does not constitute a major contribution
to the density of states (DOS).  Figure \ref{fig2}(d) shows the
partial $3d$ density of states at the Fermi level is very tiny but not
zero.  The other
occupied minority $d$ band is at -1.5 eV below the Fermi level.
Because Mn$_1(4a)$'s $d$ band is away from $E_F$ and has a small
density of states around the Fermi level, optical excitation at Mn$_1$
is weak \cite{bonfiglio2021}.

Mn$_2(4c)$ is quite different from Mn$_1(4a)$. Figure \ref{fig3}(f)
shows that its majority bands cross the Fermi level at multiple
points, have mixed $d$ characters, and are highly dispersive.  Its
$3d$-pDOS (Fig. \ref{fig3}(e)) has a larger peak at the Fermi level
than Mn$_1$, quantitatively 1.80-1.81 states/eV for the former and
0.65-0.69 states/eV for the latter. This explains why Mn$_2(4c)$ is
more optically active than Mn$_1(4a)$ \cite{bonfiglio2021}, where we
call Mn$_2(4c)$ the optical active site. In the minority channel,
Mn$_2$ has a strong admixture of orbital characters (see
Fig. \ref{fig3}(h)), and its overall density of states at the Fermi
level is also small (see Fig. \ref{fig3}(g)). We note that the minority
band structure is very similar to that of Mn$_3$Ga \cite{wurmehl2006},
and they both have a flat $d_{z^2}$ band along the $\Gamma$-X
direction.

Before we move on to ultrafast demagnetization, we must emphasize that
the band structure is not solely contributed by these two Mn
atoms. Both Ru and Ga significantly affect the magnetic properties of
Mn atoms.  Thin lines in Figs. \ref{fig2}(e) and \ref{fig2}(g) are the
Ru's $4d$ pDOS for the spin majority and minority channels,
respectively. One can see that the Ru-$d$ majority density of states
follows the Mn$_1(4a)$'s pDOS (compare Figs. \ref{fig2}(b) and (e)),
but its minority state follows the Mn$_2(4c)$'s pDOS (compare the thin
and thick lines in Fig. \ref{fig2}(g)). The split role from the same
atom is remarkable.


\subsection{Ultrafast demagnetization}

 The circles in Fig. \ref{moke}(a) are the experimental ultrafast
 demagnetization \cite{bonfiglio2021}, and consist of two
 regions. Region I is from 0 to 0.26 ps, highlighted by the red arrow
 in Fig. \ref{moke}(a), and region II starts from 0.26 ps to 5
 ps. This time separation of 0.26 ps is consistent with a prior study
 \cite{radu2011}.  In region I, a sharp decrease in spin moment is
 observed, but in region II there is a peak.  We can fit these two
 regions with the same equation, \be \frac{\Delta M(t)}{M}=A\left (
 \frac{M_{4a}{\rm e}^{-\alpha_{4a}(t-{\cal T})}-M_{4c}{\rm
     e}^{-\alpha_{4c}(t-{\cal T})}}{M_{4a}-M_{4c}}\right
 )-B, \label{demag}\ee where $t$ is the time.  $A$ is necessary, since
 without it the laser field amplitude cannot enter the equation. $B$
 determines the net amount of demagnetization.  ${\cal T}$ sets the
 characteristic time for demagnetization or remagnetization.  Since
 our spin moments are fixed by our calculation, we only have four
 fitting parameters for each region, where $\alpha_{4a(4c)}$ is the
 demagnetization rate for site $4a(4c)$.  Table \ref{tab} shows that
 in region I, $\alpha$ is site-dependent, $\alpha_{4a}=4.5$/ps and
 $\alpha_{4c}=2.8$/ps, demonstrating that the larger the spin moment
 is, the larger $\alpha$ becomes, \be \alpha=cM, {\rm or},
 \tau_M=\frac{1}{cM}, \ee where $c$ is a constant.  This equation is
 consistent with the empirical formula proposed by Koopmans and
 coworkers \cite{koopmans2005}.  From $\alpha$, we find the
 demagnetization times $\tau_M(4a)=222$ fs, and $\tau_M(4c)=357$
 fs. These intrinsic demagnetization times, called the H\"ubner times
 \cite{ourbook}, are well within the times for other transition and
 rare-earth metals: 58.9 fs (Fe), 176 fs (Ni), 363 fs [Gd($5d$)], 690
 fs [Gd($4f$)].  An extreme point will appear if $\partial \left (
 \frac{\Delta M(t)}{M}\right )/ \partial t =0$, and the second-order
 time-derivative determines whether the extreme is a maximum or
 minimum, \be \frac{\partial^2 \left (\frac{\Delta M(t)}{M}\right )
 }{\partial t^2}=A\alpha_{4c}M_{4c}{\rm e}^{-\alpha_{4c}(t-T)}
 (\alpha_{4a}-\alpha_{4c}).  \ee If $\alpha_{4a}>\alpha_{4c}$, we only
 have a minimum which explains the spin change in region I.  In region
 II, both $\alpha_{4a}$ and $\alpha_{4c}$ are reduced, but
 $\alpha_{4a}$ is reduced much more, so $\alpha_{4a}<\alpha_{4c}$,
 which corresponds to a peak in region II. Table \ref{tab} shows that
 region II has $\alpha_{4a}=0.6$/ps and $\alpha_{4a}=1.5$/ps. The
 demagnetization on the $4a$ site slows down significantly.

\subsection{Kerr rotation angle} 

Underlying ultrafast demagnetization and subsequent all-optical spin
switching is the magneto-optical property of \mrge, which is
characterized by the conductivity \cite{allen} in units of $\rm
(\Omega m)^{-1}$, \be \sigma_{\alpha\beta}(\omega)=\frac{i\hbar
  e^2}{m_e^2V}\sum_{k;m,n}\frac{f_{nk}-f_{mk}}{E_{mk}-E_{nk}}
\frac{\la nk |p_\alpha|mk\ra \la mk |p_\beta|nk\ra}
     {(\hbar\omega+i\eta)+(E_{nk}-E_{mk})}, \label{cond}\ee where
     $m_e$ is the electron mass, $V$ is the unit cell volume, $f_{nk}$
     is the Fermi distribution function, $E_{mk}$ is the band energy
     of state $|mk\ra$, $\la nk |p_\alpha|mk\ra$ is the momentum
     matrix element between states $|mk\ra$ and $|nk\ra$, and $\eta$
     is the damping parameter. The summation is over the crystal
     momentum $k$ and all the band states $|mk\ra$ and $|nk\ra$, and
     $\omega$ is the incident photon frequency. Here $\alpha$ and
     $\beta$ refer to the directions, such as the $x$ and $y$
     directions, not to be confused with the above demagnetization
     rate.  The anomalous Hall conductivity is just the off-diagonal
     term. In the limit of $\eta,\omega \rightarrow 0$, the term
     behind the summation over $k$ is the Berry curvature \be
     \Omega_{\alpha,\beta}^{k,n}=\sum_{m\ne
       n}\hbar^2\frac{(f_{mk}-f_{nk}) \la nk|v_\alpha|mk \ra \la
       mk|v_\beta|nk\ra }{(E_{mk}-E_{nk})^2}. \ee

The general expression given in Eq. \ref{cond} is better suited for
metals with partial occupation than the treatment with a separate sum
over occupied and unoccupied states \cite{wang1975,oppeneer1992},
though the latter is faster.  The intraband transition with $n=m$ is
included by replacing $(f_{nk}-f_{mk})/(E_{mk}-E_{nk})$ by its
derivative $-\partial f_{nk}/\partial E_{nk}$, which is
$-\frac{\beta/2}{\cosh\beta(E_{nk}-E_F)+1}$, without resorting to more
complicated numerics \cite{cazzaniga2010}. Here $\beta=1/(k_BT)$,
where $k_B$ is the Boltzmann constant, $T$ is the temperature, and
$E_F$ is the Fermi energy.  The Kerr effect is characterized by the
Kerr rotation $\theta$ and ellipticity $\epsilon$, in the small angle
limit and with magnetization along the $z$ axis, \be
\theta+i\epsilon=-\frac{\sigma_{xy}}{\sigma_{xx}\sqrt{1+4\pi
    i\sigma_{xx}/\omega}},\ee where $\sigma_{xx}$ must be converted to
1/s. The SI version of $\sqrt{1+4\pi i\sigma_{xx}/\omega}$ is
$\sqrt{1+ i\sigma_{xx}/(\omega\epsilon_0)}$.

Experimentally, Fleischer \et \cite{fleischer2018} measured the
magneto-optical Kerr effect for a series of Mn$_2$Ru$_x$Ga samples
with compositions $x=0.61,0.62,0.69,0.83$ and with thickness from 26
to 81 nm. Figure \ref{moke}(b) reproduces two sets of data from their
supplementary materials.  One can see that both the thickness and
composition affect the Kerr rotation angle.  The thicker sample has a
larger angle (compare dotted and long-dashed lines with
$x=0.61,0.62$), and the angle peaks between 1.6-1.9 eV.  Our
theoretical Kerr angles with three different dampings are three solid
lines with $\eta=0.8$, 0.6 and 0.4 eV from the bottom to top,
respectively.  One notices that the overall shape is similar to the
experimental data, and the main peak is also around 2 eV, slightly
higher than the experimental one, but a more direct comparison is not
possible since there are no experimental data at $x=1$. The best
agreement in terms of the Kerr angle is obtained with $\eta=0.8$ eV.
The convergence of our spectrum is tested against the mesh of
$92\times 92 \times 92$, and there is no visual difference between
this much bigger mesh and the one used in Fig. \ref{moke}(b).

We can pinpoint the origin of the main peak by removing some atoms. We
use $\eta=0.4$ eV since it gives us more structures. First, we remove
Mn$_2(4c)$, without changing the lattice structure and the rest of
atoms, so we have Mn$(4a)$RuGa.  The solid line in Fig. \ref{moke}(c)
shows that the Kerr rotation angle for the new Mn$(4a)$RuGa is very
different from the one in Fig. \ref{moke}(b), highlighting the fact
that Mn$_2(4c)$, not Mn$_1(4a)$, contributes significantly to the
overall signal. To verify this, we remove Mn$_1(4a)$ but keep
Mn$_2(4c)$. The red long-dashed line shows clearly that the overall
shape is well reproduced, but the Kerr angle is larger. This concludes
that Mn$_2(4c)$ is optically active and plays a decisive role in the
magneto-optical response as OAS, consistent with the experiment
\cite{bonfiglio2021}, but the role of Ru and Ga should not be
underestimated.  Magnetically, they are silent and do not contribute
to the spin moment significantly, but when we remove Ru, the spectrum
changes completely (see the dotted line in Fig. \ref{moke}(c). The
same thing happens to the removal of Ga atom. This reveals the
significant contributions of Ru and Ga to the optical response of
\mrge.  {\clr The discovery of two sites (spin anchor site and optical
  active site) found here has some resemblance to the laser-induced
  intersite spin transfer \cite{dewhurst2018}. In their system,
  Dewhurst \et found that the spins of two Mn atoms are aligned and
  coupled ferromagnetically, not antiferromagnetically coupled as
  found here. Additional calculations are necessary since their
  materials are not \mrge.  Mentink \et \cite{mentink2012} proposed a
  two-sublattice spin model where AOS is realized through the angular
  momentum exchange between sublattices.  But their did not reveal the
  different roles played by two spin sublattices. Our mechanism makes
  a clear distinction between two sublattices, and thus ensures that
  two sublattices do not compete optically and magnetically.}

\section{Conclusion}

\begin{figure}
  \includegraphics[angle=0,width=1\columnwidth]{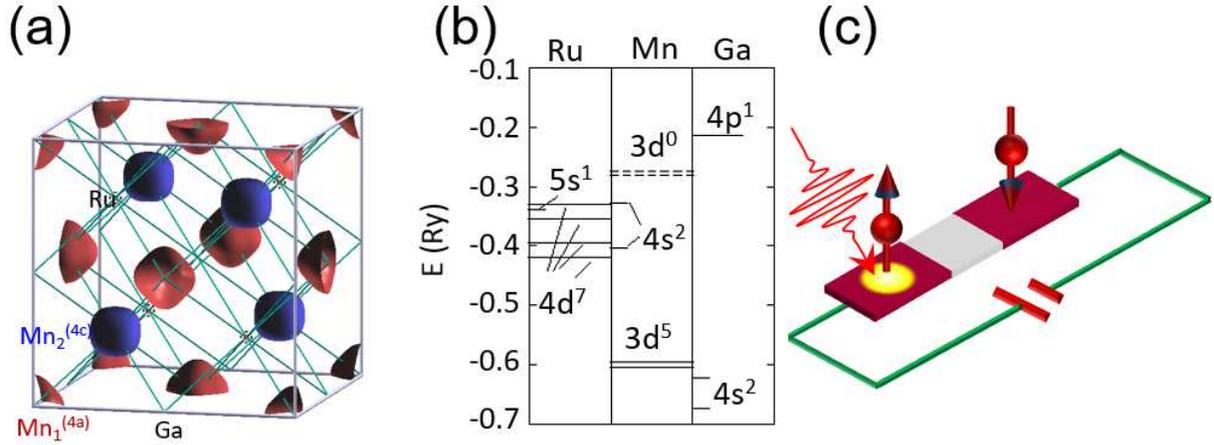}
\caption{(a) Structure of \mrg and the spatial spin densities on the
  Mn$_1(4a)$ (red, positive) and Mn$_2(4c)$ (blue, negative). The Ru
  and Ga atoms have a small spin density. (b) Atomic energy levels of
  Ru, Mn and Ga. The energy splitting is due to the spin-orbit
  coupling in atoms.  (c) Our proposed device has a junction structure
  and consists of three layers of the same Mn$_2$Ru$_x$Ga, but with
  different composition $x$. The layer on the left is an optically
  active layer, the middle is the spin filter, and the right layer is
  a spin reference layer. The magnetoresistance is controlled by
  light.  }
\label{fig01}
\end{figure}

\begin{figure}
  \includegraphics[angle=0,width=0.7\columnwidth]{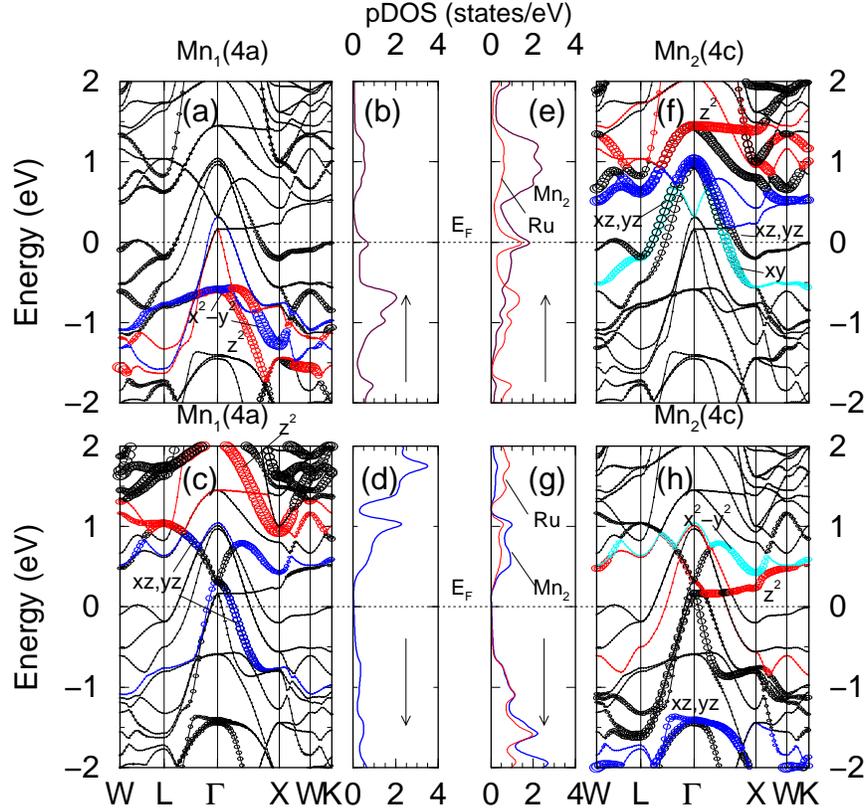}
\caption{(a) and (c) Orbital-resolved band structure with the $3d$
  state characters for the Mn$_1(4a)$ spin-majority and spin-minority
  channels, respectively.  (b) and (d) Partial density of states for
  the Mn$_1$ spin-majority and spin-minority channels, respectively.
  Bands with clear orbital characters are denoted by their orbitals.
  The Fermi level is set at 0 eV (horizontal dashed line).  (f) and
  (h) Band structure with the $3d$ state characters for the
  Mn$_2(4c)$ spin-majority and spin-minority channels, respectively.
  (e) and (g) Partial density of states for the Mn$_2$'s $3d$ (thick
  lines) and Ru's $4d$ (thin lines) spin-majority and spin-minority,
  respectively.  }
\label{fig2}
\label{fig3}
\end{figure}

\begin{figure}
  \includegraphics[angle=0,width=0.8\columnwidth]{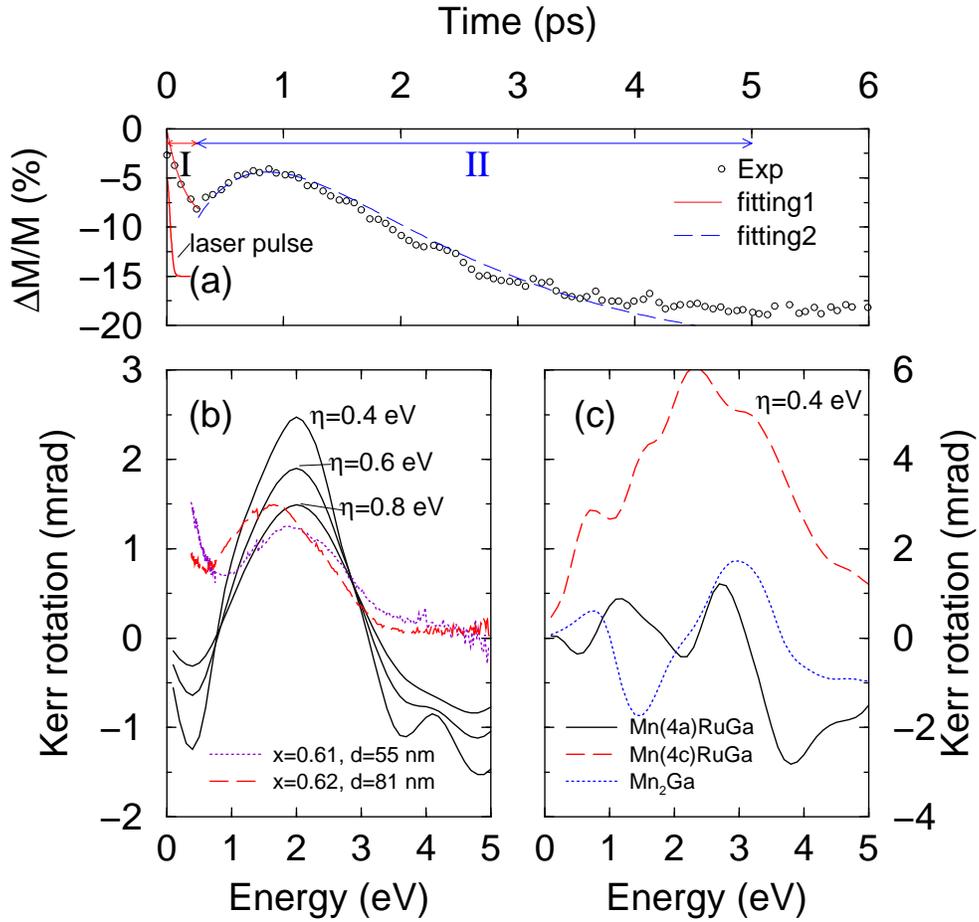}
  \caption{(a) Experimental demagnetization fitted by Eq. \ref{demag},
    with two sets of fitting parameters given in Table \ref{tab},
    provides a crucial insight that demagnetization rates at two Mn
    spin-sublattices change between region I (between 0 to 0.26 ps)
    and region II (between 0.26 ps to 5 ps). The experimental data are
    extracted from Ref. \cite{bonfiglio2021}. {\clr The thick red
      curve is the laser pulse of duration 40 fs.}  (b) The
    experimental (dotted and dashed lines from
    Ref. \cite{fleischer2018}) and our theoretical Kerr rotation
    angles. The three solid lines are our theoretical results with
    three different dampings $\eta=0.4,0.6,0.8$ eV. (c)
    Element-resolved Kerr rotation angles when Mn$_2(4c)$ (solid
    line), or Mn$_1(4a)$ (dashed line), or Ru (dotted line) is removed
    separately. $\eta=0.4$ eV is used.  }
\label{moke}
  \end{figure}

Through \mrge, our state-of-the-art first-principles density
functional calculation establishes two concepts: the spin anchor site
and the optical active site as the key to AOS in ferrimagnets. The
formation of SAS and OAS in \mrg is accomplished by weakly magnetic Ru
and Ga atoms. In GdFeCo \cite{stanciu2007} Gd is SAS while Fe is OAS.
Switching starts with OAS \cite{radu2011}; because the ferrimagnetic
coupling between SAS and OAS is frustrated and has a lower potential
barrier to overcome if the spin moment is smaller \cite{prb17}, SAS is
dragged into the opposite direction by OAS through the spin torque
$J{\bf S}_i\times {\bf S}_{j}$ \cite{epl16,aip20}, to realize
all-optical spin switching. Because the Heusler compounds have
excellent tunability
\cite{liu2006,betto2016,wollmann2017,song2017,zhang2018a,fleischer2018},
the future research can investigate the effect of the spin moments at
SAS and OAS on the spin switchability \cite{vahaplar2012}.  We
envision an integrated device based on \mrg as illustrated in
Fig. \ref{fig01}(c). All three parts of the device are made of the
same materials but with different concentration $x$. In the middle,
$x$ is close to 0.6, so we have a full-compensated half-metal, while
at two ends, $x$ is close to 1, whose spin is designed to be optically
switched. This forms an ideal magnetic tunnel junction that a light
pulse can activate. A future experimental test is necessary.  {\clr We
  should add that experimentally, concentrations of both Mn and Ru can
  be already tuned in several different experimental groups.
  Chatterjee \et \cite{chatterjee2021} were able to adopt two
  concentrations $x=0.2,0.5$ in Mn$_{2-x}$Ru$_{1+x}$, while Siewierska
  \et \cite{siewierska2021} were able to independently change $x$ and
  $y$ in $\rm Mn_yRu_xGa$ films. In fact, Banerjee \et
  \cite{banerjee2020} already used 13 samples with $x=0.5$ up to 1.0
  in $\rm Mn_2Ru_xGa$.}

\acknowledgments

The authors appreciate the numerous communications with Dr. K. Rode
(Dublin) and Dr. K. Fleischer (Dublin). Dr. Fleischer provided the
original experimental data in text form, which is very convenient to
plot, with a small correction to the thickness of their samples.
G.P.Z. and Y.H.B.  were supported by the U.S. Department of Energy
under Contract No. DE-FG02-06ER46304. Part of the work was done on
Indiana State University's high performance Quantum and Obsidian
clusters.  The research used resources of the National Energy Research
Scientific Computing Center, which is supported by the Office of
Science of the U.S. Department of Energy under Contract
No. DE-AC02-05CH11231. M.S.S. was supported by the National Science
Foundation of China under grant No. 11874189.

$^*$ guo-ping.zhang@outlook.com.
 https://orcid.org/0000-0002-1792-2701

{\clr

\begin{table}
\caption{Summary of the space group symmetries for X$_2$YZ used in the
  literature. Two X atoms are denoted as X$_1$ and X$_2$, and
  ``share'' means that they share the same positions. For \mrge, X$_1$
  is Mn$_1$, X$_2$ is Mn$_2$, while Y is Ru and Z is Ga.  The
  full-Heusler compound has L2$_1$ symmetry, the half-Heusler one has
  $C1_b$ symmetry, and the inverse Heusler compound has $XA$ symmetry.
}
\begin{tabular}{l|l||ll|ll|l}
\hline\hline
Group symmetry & Prototype  &X$_1$ & X$_2$ &Y & Z& Ref.\\
\hline
L2$_1$ (No. 225, $Fm\bar{3}m$) & Cu$_2$MnAl &
$8c(\frac{1}{4},\frac{1}{4},\frac{1}{4})$&share &$4b(\frac{1}{2},\frac{1}{2},\frac{1}{2})$& $4a(0,0,0)$& \cite{wollmann2014}\\
L2$_1$ (No. 225, $Fm\bar{3}m$) & &
$(0,0,0)$&$(\frac{1}{2},\frac{1}{2},\frac{1}{2})$
&$(\frac{1}{4},\frac{1}{4},\frac{1}{4})$&
$(\frac{3}{4},\frac{3}{4},\frac{3}{4})$& \cite{galanakis2002a}\\
L2$_1$ (No. 225, $Fm\bar{3}m$) & Cu$_2$MnAl 
&$8c(\frac{1}{4},\frac{1}{4},\frac{1}{4})$&
$(\frac{3}{4},\frac{3}{4},\frac{3}{4})$&
$4a(0,0,0)$&$4b(\frac{1}{2},\frac{1}{2},\frac{1}{2})$
& \cite{pearson2}\\
\hline
$C1_b$ (No. 216,  $F\bar{4}3m$) &MgAgAs &  $4a(0,0,0)$& vacant &
$4b(\frac{1}{2},\frac{1}{2},\frac{1}{2})$ &
$4c(\frac{1}{4},\frac{1}{4},\frac{1}{4})$ & \cite{pearson2}\\
\hline
$XA$ (No. 216, $F\bar{4}3m$) & Li$_2$AgSb  &$4d(\frac{1}{4},\frac{1}{4},\frac{1}{4})$&$4b(\frac{1}{2},\frac{1}{2},\frac{1}{2})$
&$4c(\frac{3}{4},\frac{3}{4},\frac{3}{4})$& $4a(0,0,0)$&
\cite{wollmann2014}\\
$XA$ (No. 216, $F\bar{4}3m$) &  &
$4a(0,0,0)$&$4c(\frac{3}{4},\frac{3}{4},\frac{3}{4})$&$4b(\frac{1}{2},\frac{1}{2},\frac{1}{2})$
&$4d(\frac{1}{4},\frac{1}{4},\frac{1}{4})$ &
\cite{galanakis2014}\\
No. 216 &  & $4a(0,0,0)$& $4c(\frac{1}{4},\frac{1}{4},\frac{1}{4})$&$4d(\frac{3}{4},\frac{3}{4},\frac{3}{4})$
&$4b(\frac{1}{2},\frac{1}{2},\frac{1}{2})$
&\cite{hahn}\\
\hline 
 \end{tabular}
\label{table0}
\end{table}

}

\begin{table}

\caption{Spin and orbital moments of Mn$_1$, Mn$_2$, Ru and Ga. The
  electron populations in their majority and minority $3d$ states are
  listed as $n_{3d,\uparrow}$ and $n_{3d,\downarrow}$. The
  demagnetization rate in regions I and II is denoted as $\alpha^I$
  and $\alpha^{II}$, respectively. In two artificial structures, $\rm
  Mn_2Ru$ and  $\rm Mn_2Ga$, only the spin moments are given.  
}

\begin{tabular}{ccccccc|cc}
\hline\hline
Element & $M_s(\ub)$ & $M_o(\ub)$
&$n_{3d\uparrow}$&$n_{3d\downarrow}$&$\alpha^I$(1/ps)&$\alpha^{II}$(1/ps)
&$M_s(\ub)$ 
&$M_s(\ub)$\\
        & (\mrge)           & (\mrge)          
&                &                  &                &
&
(Mn$_2$Ru)
&        (Mn$_2$Ga)
\\
\hline
Mn$_1(4a)$& 3.17&0.025 &4.09 &1.01 &4.5 &0.6 &2.80 & 3.35\\
Mn$_2(4c)$&-2.31&-0.046&1.44 &3.72 &2.8 &1.5&-3.64& -3.27\\
Ru$(4d)$ & 0.076 & -0.035 &&&&& -0.39 & NA\\
Ga$(4b)$ & 0.032 & -0.000 &&&&& NA    &-0.04\\
\hline\hline
\\
\end{tabular}
\label{tab}
\end{table}

\end{document}